\documentstyle[twocolumn,aps,epsf]{revtex}    

\begin{document}
\draft                             

\title{Experimental evidence for the formation of stripe phases in Si/SiGe}

\author{U.~Zeitler,$^1$\cite{email} 
H.~W.~Schumacher,$^1$\cite{present}
A.~G.~M.~Jansen,$^2$ and R.~J.~Haug$^1$}
\address{
$^1$Institut f\"ur Festk\"orperphysik, Universit\"at Hannover, 
Appelstra{\ss}e 2, 30167 Hannover, Germany\\
$^2$Grenoble High Magnetic Field Laboratory,
Max-Planck-Institut f\"ur Festk\"orperforschung and 
Centre National de la Recherche Scientifique,
B.P.~166, 38042 Grenoble Cedex 09, France}

\date{\today}        
\maketitle

\begin{abstract}

We observe pronounced transport anisotropies in magneto-transport 
experiments performed in the two-dimensional electron system of a 
Si/SiGe heterostructure. They occur when an in-plane field is used to
tune two Landau levels with opposite spin to energetic coincidence.
The observed anisotropies disappear drastically for temperatures above 1 K.
We propose that our experimental findings may be caused by the formation 
of a unidirectional stripe phase oriented perpendicular to the in-plane field.

\end{abstract}


\narrowtext


Two-dimensional electron systems (2DESs) in high magnetic 
fields show many interesting fundamental effects.
Widely known is the quantization of the 
energy spectrum into discrete Landau levels (LLs).
For high magnetic fields the electron-electron interaction
becomes important and novel states emerge.
One of the most prominent of such features is the fractional 
quantum Hall effect (FQHE)~\cite{FQHE} where new collective quasi-particles 
for electrons in the lowest two LLs appear.

For higher LLs it has been proposed that an
ordered charge density wave forms an energetically lower 
lying state compared to the FQHE states~\cite{Fogler96,Moessner96}.
In the situation when a higher LL
is half filled the electrons arrange in stripes where the LL considered
is either totally full or totally empty.
Experimental evidence for the existence of such a stripe phase
at half LL filling
was found recently by several groups in 2DESs of very high 
mobility GaAs/AlGaAs 
heterostructures~\cite{Lilly99-1,Du99,Pan99,Lilly99-2}.
With an additional magnetic field $B_{ip}$ parallel to the 2DES 
the orientation of the stripes 
can be tuned to the direction perpendicular
to $B_{ip}$~\cite{Pan99,Lilly99-2,Jungwirth99}.


In this Letter we report on magneto-transport experiments 
in the 2DES of a Si/SiGe heterostructure. Adding an in-plane field $B_{ip}$,
with the normal field component left constant,  
two neighboring LLs with  
opposite spin can be tuned to half filling simultaneously. 
We will show that huge maxima in the Shubnikov-de Haas (SdH) oscillations 
appear if the current direction $I$ is oriented along $B_{ip}$.
Such an enhancement of the SdH-maxima is not observed when $I$ is
oriented perpendicular to $B_{ip}$. We will propose the formation
of a unidirectional ''stripe phase'' to explain the experimentally observed
huge transport anisotropies and possible physical origins of 
the stripe formation will be discussed.
    


Our sample is a Si/SiGe heterostructure  with a 25-nm thick 
strained Si channel embedded between two Si$_{0.7}$Ge$_{0.3}$ 
barriers~\cite{SiGe}.
The electrons are provided by
doping the top barrier with Sb starting 12 nm away from the Si channel.
The resulting band-structure leads to a high mobility 
2DES formed in a triangular potential at the 
heterojunction between the Si channel and the top SiGe barrier
(electron concentration $n = 7.2 \times 10^{15}$~m$^{-2}$, mobility
$\mu = 20$~m$^{2}$/Vs). 
In order to perform transport experiments 
a 100-$\mu$m wide Hall bar was patterned on the sample along
the [110] direction.


In a magnetic field the energy level structure 
of the 2DES consists of discrete LLs 
at energies $E_N = (N+1/2) (\hbar e B_n / m^* )$,
where $N =0,1,2,...$ is the LL index,
$B_n$ is the field component perpendicular
to the 2DES and $m^* = 0.19 m_e$ is the effective
electron mass in Si. 
Each LL is split into two spin levels, 
$E_{N,s} = E_N \pm \frac{1}{2} g^* \mu_B B$. Here
$s =~{\tiny \uparrow/\downarrow}$ denotes the spin orientation,
$g^*$ is the effective Land\'e factor, and $B$ 
is the {\sl total} magnetic field.
Additionally, each spin level consists 
of two distinct valleys~\cite{AFS} resolved as individual levels
in transport experiments in Si/SiGe heterostructures \cite{Weitz,Koester}. 



With a magnetic field oriented perpendicular to the 2DES 
the longitudinal resistance $\rho_{xx}$
displays Shubnikov-de Haas (SdH) oscillations, see Fig.~1a, bottom
trace, where 
$\rho_{xx}$ is plotted as a function of the
LL filling factor $\nu = h n / eB_n$, where $B_n$ is the normal 
field component with respect to the orientation of the 2DES.
The SdH-oscillations are characterized by pronounced minima at 
filling factors $\nu= 4(N+1) = 4,8,12,16,$ etc. Here
the Fermi level $E_F$ is situated inside the gap between two
neighboring LLs $N$ and $N+1$. 
Additional minima occur at $\nu = 2+4N = 2,6,10,14,18,$ etc.~when $E_F$
is positioned between the two spin-split sub-levels inside a LL $N$. 
For the two lowest LLs $N=0$ and $N=1$ also
the valley splitting is resolved visible as minima at 
$\nu=3,5,7, {\rm and}~9$.

 
The experiments described in this Letter use tilted magnetic fields.
Before analyzing the results  we will shortly introduce into the 
coincidence technique used in such experiments~\cite{Fang}.
More details concerning the specific experiment can be found
in \cite{wir}. The main idea of this technique lies in the possibility
of modifying the LL structure of a 2DES 
in a perpendicular magnetic field $B_n$ 
by adding  an additional in-plane field $B_{ip}$. 
For simplicity we will first not consider valley splitting in the following
description.

In our Si/SiGe structure the spin splitting without in-plane field 
amounts to approximately one third of the LL 
splitting.
Adding $B_{ip}$ with $B_n$ constant leaves the energetic position
of the center of a LL constant.  
However, the spin splitting $\Delta E_Z= g^* \mu_B B$  
between two spin-split levels ($N,\uparrow$) and ($N,\downarrow$),
depends on the total magnetic field $B = (B_n^2 + B_{ip}^2)^{1/2}$, 
and thus increases when adding an in-plane field.

The resulting energy level structure is illustrated in Fig.~1b, supposing
a constant effective g-factor $g^* = 3.5$ corresponding to the
situation where the spin splitting equals one third of the LL splitting
in a perpendicular field. 
In fact, $g^*$ in Si/SiGe heterostructures is also
dependent on the LL filling as well as on the strength of the
in-plane field. For illustration purposes we do not take into account
this complication, see Refs.~\onlinecite{Weitz,Koester,wir} 
for more details.

The LL energy $\epsilon_{N,s}$ (in units of $\hbar e B_n / m^*$,
bottom axis) is plotted as a function of the total field (normalized to
the normal field component $B_n$, left axis).
Each spin-split level consists of two valley sub-levels 
marked as two parallel lines in the figure.
An increase of the total field while leaving $B_n$ constant leads to a relative
increase of the spin splitting $\Delta E_Z$ compared to the LL splitting
$\hbar \omega_c$.
As soon as $\Delta E_Z$
equals $\hbar \omega_c$ two neighboring LLs with
opposite spin, ($N+1,\downarrow)$ and ($N,\uparrow)$,
are situated at the same energy, they coincide.
As a consequence, the Fermi energy at
filling factors $\nu= 4 (N+1) = 4,8,12,16,20,$ etc.~is 
no more situated in a gap but inside these degenerate levels.
The pronounced minima in $\rho_{xx}$ at filling factors $\nu=4 (N+1)$
change into maxima. This coincidence is found experimentally
at a tilt angle $\vartheta = 72.4^o$, see Fig.~1a, where  
SdH maxima are found at filling factors $\nu =8,12,16,$ and 20.  
In the data presented in the figure no coincidence maximum can be seen
at $\nu=4$. 
In fact a very pronounced SdH maximum only appears in a very narrow 
angle range around $\vartheta_1^* = 70^o$. 
This observation will be analyzed in more detail below.

 
Higher-order coincidences occur when the spin splitting equals an integer
multiple of the LL splitting. 
In particular the second-order coincidence is found experimentally at
$\vartheta = 80.5^o$ when $\Delta E_Z = 2 \hbar \omega_c$. 
Now the ($N+2,\downarrow$) and the ($N,\uparrow$) levels coincide
at filling factors $\nu= 2 + 4 (N+1)$ etc.~and 
maxima appear at $\nu = 10,14,16,$ and 22 in Fig.~1a.
Again the expected coincidence at $\nu=6$ is not visible in the
data presented, see below.


As already stated above, the first-order coincidence at $\nu=4$ 
was not observed in the traces
in Fig.~1a. Therefore, we will analyze this range in more detail in
the following. In Figs.~2a and 2b a color contour plot of $\rho_{xx}$
is shown when moving through the coincidence, with some selected traces
in Figs.~2c and 2d. 
 
The most striking feature is the appearance of a strongly pronounced
transport anisotropy as a function of the
orientation of the in-plane field measured in the {\sl same} Hall bar. 
This anisotropy has been reproduced experimentally in several 
cool-down cycles of the sample and on different voltage 
contacts of the Hall bar.
 
Before the coincidence ($\vartheta = 68.5^o$) the SdH-oscillations of
$\rho_{xx}$ look similar for both orientations
of $B_{ip}$ with respect to the current direction, see bottom
traces in Figs.~2c and 2d. The slight differences are most probably
due to slightly different 2DES properties originating
from different cool-down cycles. 
When moving towards the coincidence by increasing $\vartheta$
drastic differences start to
appear.  For $B_{ip} \parallel I$ a huge maximum 
in $\rho_{xx}$ develops reaching peak values of more than 13~k$\Omega$ at 
$\vartheta = 69.98^o$.

Also for $B_{ip} \perp I$ the coincidence shows its presence
by the disappearance of the $\nu = 4$ minimum, 
but no unusual enhancement of $\rho_{xx}$ is observed. 
In contrast, the magnitude of the SdH maximum inside 
the coincidence is with less than 1~k$\Omega$ even lower than
the typical peak values of $\rho_{xx}$ outside the coincidence.

Similar huge transport anisotropies with a strong enhancement
of $\rho_{xx}$ are observed when the spin-up
level of the lowest LL is coinciding
with a higher LL with opposite spin. Experimentally
we find this behavior for the second-order coincidence at $\nu=6$,
where the $N=0$ and the $N=2$ LLs are coinciding,
and indications for it for the third-order coincidence at $\nu=8$~\cite{RemK3}.
At all these positions huge in-plane fields $B_{ip} > 20$~T are present.

As already shown in Fig.~1 no spectacular effects occur
when only higher LLs are involved in the coincidence
where the parallel field component is comparably lower.
This statement also remains true for third and higher-order coincidences 
not shown in the figure.

In order to further investigate the properties 
of the electron system for the first-order coincidence at $\nu=4$
we have performed temperature dependent experiments for the two
orientations of the in-plane fields shown
in Fig.~3. In the insets the temperature dependence of the dominant
SdH peak in the center of the coincidence is displayed.
The strongly enhanced maxima in  $\rho_{xx}$ for the first-order coincidence
at $\nu=4$ with $B_{ip} \parallel I$  disappears
for temperatures larger than 1~K.  
With $B_{ip} \perp I$ the suppression of $\rho_{xx}$ weakens in the 
same temperature range.
Now $\rho_{xx}$ approximately doubles its value when the temperature
is increased from 0.45 K to above 1 K, see Fig.~3b.

In order to explain our experimental findings 
we propose that during the coincidence, where the 
huge anisotropy appears, a unidirectional stripe phase 
oriented perpendicular to the in-plane field is formed.
In such a picture transport along the stripes would be facilitated 
and transport across the stripes would be obstructed. 


We suggest that the effects observed are 
caused by the successive depopulation
of an initially totally filled LL ($N,\uparrow$)
with the simultaneous filling of the initially empty LL 
($N+1,\downarrow$).
In the center of the coincidence  
two charge-degenerate LLs are half-filled.
The particularity of this half filling points to possible  
similarities with recent experiments concerning a {\sl single half-filled}
higher LL in very high mobility GaAs/GaAlAs structures 
\cite{Lilly99-1,Du99,Pan99,Lilly99-2}.

Considering that the mobility of these GaAs-based 2DESs is nearly
two orders of magnitude higher than in our Si/SiGe structures
it is at first sight rather astonishing that we are able to observe
any stripe phases at all.
However, the additional valley splitting in Si/SiGe structures 
can lead to a much more complex phase diagram as compared to GaAs.
In this respect it is interesting to state that the typical size
of the valley splitting is comparable to the disorder broadening
of an individual valley- and spin-split LL.
Therefore, the presence of two valleys may well stabilize a possibly 
existing stripe phase with respect to disorder.
 
Without considering valley splitting in detail, a charge-homogeneous 
spin-density phase was shown to be unstable with respect to a first 
order phase transition into a ferromagnetic state with one completely 
filled LL \cite{Giuliani}. Such a transition was also
found experimentally in the 2DES of a GaInAs/InP heterostructure~\cite{Koch}.
However, this result does not exclude a stable charge-inhomogeneous
state. Since the spin and orbital degrees of freedom are rigidly
coupled to each other, both charge-density wave and 
spin-density wave order parameters are finite in
this type of ground state. Preliminary theoretical considerations
suggest that the charge-inhomogeneous Hartree-Fock state in 
a half-filled higher LL may survive the addition of 
electrons from the half-filled LL below \cite{Ferdi}.

From the temperature dependence shown in Fig.~3 
we can estimate the typical correlation energy for our
proposed stripe phase to be on the order of 0.1~meV. 
In this respect a stripe phase
would be energetically favorable if the energy separation of the two Landau
levels involved in the coincidence is smaller than the correlation
energy. At low temperatures where the stripes are formed transport 
across the stripes is obstructed and as a consequence the resistance 
increases drastically with decreasing temperature as soon as the stripe phase
starts to form. Transport along the stripes would be facilitated compared
to a homogeneous electron distribution, the resistance decreases 
with decreasing temperature. 

The correlation energy for a stripe formation can be deduced 
independently from the
narrow angle range  where the coincidence between the two levels 
($0,\uparrow$) and ($1,\downarrow$) appears, 
approximately $\pm 0.3^o$ around $\vartheta_1^* = 70^o$.
At a constant filling factor $\nu=4$
(corresponding to a normal field $B_n = 7.5$~T ) 
the energy separation of the two levels,
$\Delta E =  E_{0,\uparrow} - E_{1,\downarrow}$  
changes from $-0.07~$meV to $+0.07~$meV 
in a single particle picture. Here 
$\Delta E = g^* \mu_B B_n ( 1 / \cos \vartheta  - 1 / \cos \vartheta_1^*)$
is defined by the relative change of the Zeeman energy of the two levels.
$g^* = 2 (m/m^*)  \cos \vartheta_1^* = 3.6$ is the effective g-factor
at $\nu=4$ deduced from the position of the first-order coincidence at
$\vartheta_1^* = 70^o$. In other words correlations between the levels
become important as soon as their energetic separation gets below the
correlation energy.

Since the energy gain for forming our proposed stripe
phase is of the order of the typical valley splitting\cite{Weitz,Koester}
it is worthwhile speculating that the stripes are due do a redistribution
of electrons between different pockets in $k$-space and the
interaction between these valleys during the coincidence.
Without any doubt a detailed theoretical consideration of the
complex energy-level structure including valley splitting is necessary
to indicate more clearly the existence of stripe phases 
in Si/SiGe heterostructures. 
 
The proposed stripe formation might be supported by a
geometrical modulation of the Si quantum well and the adjacent SiGe
barriers. The linearly graded relaxed buffer is known to relax by long
misfit dislocations distributed over the whole thickness of the graded
buffer part~\cite{Schaeffler92}.
Each dislocation
creates a double atomic height step on the surface, dislocation
multiplication can lead to a pile-up of these surface steps. This is the
origin of the cross-hatch surface morphology, which is oriented along the [110]
directions, as are the misfit dislocations. Preferential growth on surface
steps smears out the surface steps, which leads to a smoothly varying
modulation of $\pm 2~$nm with an  average period 
of 1.3~$\mu$m for the cross hatch (measured with an atomic-force microscope
on the surface of the heterostructure).

This surface morphology can be viewed as a slight 
modulation of the orientation of the 2DES.
A strong in-plane field will then lead to a modulated tilting angle
along $B_{ip}$.
As a consequence stripes perpendicular to 
$B_{ip}$ will form where the energetically lower lying LL is
either the ($N,\uparrow$) level or the ($N+1,\downarrow$) level. 

As an example we regard again the first-order coincidence at $\nu=4$
at an average angle $\vartheta_1^* = 70^o$, see Fig.~2.
Due to the modulation of $\vartheta$, the occupation of the two LLs
involved in the coincidence is modulated along $B_{ip}$. 
Already a modulation amplitude of less than $0.6^o$ would be sufficient
to form stripes with alternating filling of the two LLs. 
Due to the cross-hatch pattern in the dislocation this stripe formation
is most pronounced if the $B_{ip}$ is oriented along [110] or [1$\bar{1}$0],
respectively, the two orientations chosen in the experiment.

In conclusion we have observed strong anisotropies induced by an
in-plane field in the magneto-transport properties of coinciding
Landau levels in the 2DES of a Si/SiGe heterostructure. 
We propose that they are caused by the formation of a unidirectional stripe
phase formed by electrons from two Landau levels with opposite spin.
From temperature dependent experiments we deduced a typical correlation
energy on the order of 0.1 meV for the formation of the stripe phase.


We would like to thank F.~Sch\"affler for providing 
the sample and for helpful discussions.
We acknowledge F.~Evers, V.~I.~Fal'ko and D.~G.~Polyakov 
for illuminating discussions, and I.~Hapke-Wurst, U.~F.~Keyser,
A.~Nauen and J.~Regul for experimental assistance.
Part of this work was supported by the TMR Programme of the 
European  Union under contract no.~ERBFMGECT950077.

\vspace*{-1.5em}                  

\begin{figure}[t]
\centerline{\epsfxsize=4.5cm              
\epsfbox{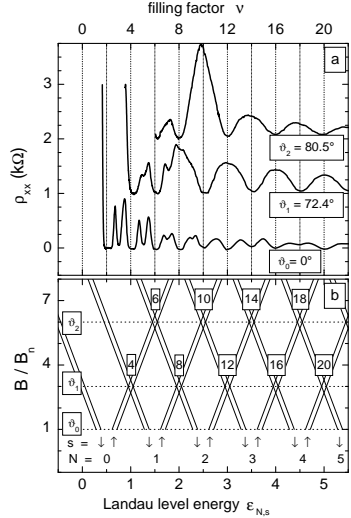}}                     
\vspace*{1em}
\caption{(a) Resistivity $\rho_{xx}$ as a function of the filling 
factor $\nu$ for tilt angles $\vartheta_0 = 0^o$, 
$\vartheta_1 =72.4^o$  (first-order coincidence) and 
$\vartheta = 80.5^o$ (second-order coincidence).\\
(b) Landau-level structure at constant normal field $B_n$ as a function
of the total magnetic field $B$ (in units of $B_n$).
The numbers indicate the even integer filling factors where 
coincidences occur. 
At the horizontal line marked $\vartheta_0$ the magnetic field
is oriented perpendicular to the 2DES, the lines marked
$\vartheta_1$ and $\vartheta_2$ 
sketch the angle positions where the first- and the second-order 
coincidences are found.}
\label{Rxx}
\end{figure}

\begin{figure}
\centerline{\epsfxsize=7.5cm         
\epsfbox{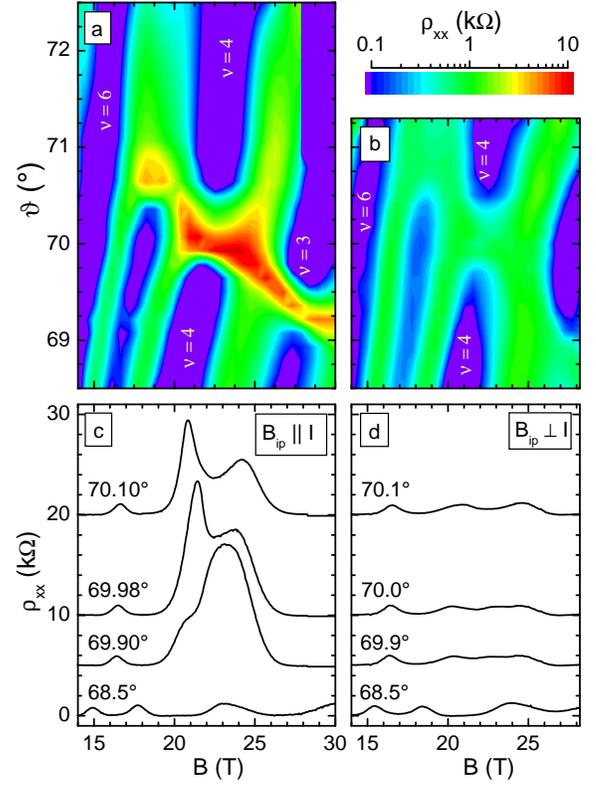}}                  
\vspace*{1em} 
\caption{{\sl top:} Color contour plot of $\rho_{xx}$ in the first 
order coincidence
around $\nu=4$ on a logarithmic scale from 100~$\Omega$ (blue) via 
1~k$\Omega$ (green) to 10~k$\Omega$ (red). 
Values lower than 100~$\Omega$ and missing values are violet,
values higher than 100~k$\Omega$ are dark red. In (a) the in-plane field
$B_{ip}$ is oriented along the current in (b) $B_{ip}$ is perpendicular 
to the current.\\
{\sl bottom:} Selected traces of $\rho_{xx}$ in the coincidence 
for the two orientations of the in-plane field. The curves are shifted for
clarity.}
\label{3D}
\end{figure}

\begin{figure}
\centerline{\epsfxsize=7cm           
\epsfbox{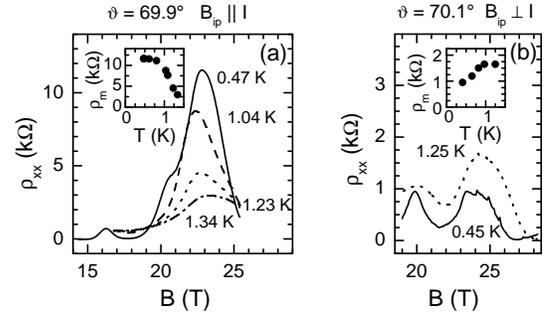}}                  
\vspace*{1em} 
\caption{Temperature dependence of $\rho_{xx}$ in the first-order coincidence
at $\nu=4$. In (a) the in-plane field is parallel to the current
in (b) $B_{ip}$ is perpendicular to $I$. The insets show the temperature
dependence of the dominant SdH for both field orientations.}  
\label{Tabh}
\end{figure}


\begin{references}
\bibitem{RemK3}
The magnetic field available (30~T) did not allow to measure
the $\rho_{xx}$-anisotropy along the entire third-order coincidence
at $\nu=8$. However, a strong increase of $\rho_{xx}$ 
(up to $\rho_{xx} = 4~$k$\Omega$) with $B_{ip} \parallel I$ 
is observed when approaching the coincidence.
\end{references}
\end{document}